\documentclass[runningheads]{llncs}
\usepackage{float}
\restylefloat{table}
\usepackage{graphicx}
\usepackage{tabularx} 
\usepackage{algorithm}
\usepackage{algpseudocode}
\usepackage{wrapfig}
\usepackage{csvsimple}
\usepackage[sorting=none]{biblatex}
\begin{document}
\title{%
NFT Wash Trading\\
\large Quantifying suspicious behaviour in NFT markets}

\titlerunning{NFT Wash Trading}

\author{Victor von Wachter\inst{1} \and
Johannes Rude Jensen\inst{1,2} \and
Ferdinand Regner\inst{3} \and
Omri Ross\inst{1,2}}
\authorrunning{von Wachter et al.}
% If there are more than two authors, ‘et al.’ is used.
%
\institute{University of Copenhagen, Denmark \and
eToroX Labs, Denmark \and University of Vienna, Austria}
\maketitle % typeset the header of the contribution
\begin{abstract}
The smart contract-based markets for non-fungible tokens (NFTs) on the Ethereum blockchain have seen tremendous growth in 2021, with trading volumes peaking at \$3.5b in September 2021. This dramatic surge has led to industry observers questioning the authenticity of on-chain volumes, given the absence of identity requirements and the ease with which agents can control multiple addresses. We examine potentially illicit trading patterns in the NFT markets from January 2018 to mid-November 2021, gathering data from the 52 largest collections by volume. Our findings indicate that within our sample 3.93\% of addresses, processing a total of 2.04\% of sale transactions, trigger suspicions of market abuse. Flagged transactions contaminate nearly all collections and may have inflated the authentic trading volumes by as much as \$149,5m for the period. Most flagged transaction patterns alternate between a few addresses, indicating a predisposition for manual trading. We submit that the results presented here may serve as a viable lower bound estimate for NFT wash trading on Ethereum. Even so, we argue that wash trading may be less common than what industry observers have previously estimated. We contribute to the emerging discourse on the identification and deterrence of market abuse in the cryptocurrency markets. 

\keywords{DeFi \and NFT \and Blockchain \and Wash trading \and Graph analysis}
\end{abstract}

%%%%%%%%%%%%%%%%%%%%%%%%%%%%%%%%%%%%%%%%%%%%%%%%%%%%%%%%%%%%%%%%%%%%%%%%%%%%%%%%

\section{Introduction}
A non-fungible token (NFT) is a unique digital representation of a digital or physical asset. While the NFT standard is used widely to designate ownership of artefacts such as domain name registrations or concentrated liquidity positions in constant function market makers (CFMM) \cite{uniswap_2021}, the arguably most recognized use of the NFT standard is within the representation and trade of digital art and collectibles. Here, the NFT is typically used to represent the ownership of a digital image externally stored, either on a server or, more commonly, on censorship resistant distributed file systems such as the Interplanetary File System. A basic NFT standard such as the ERC721 \cite{entriken_2018} typically denotes an interface implementing the ability to own, transfer and trade the NFT. Standardization led to the emergence of NFT markets, facilitating primary and secondary trading, the presently most dominant of which is OpenSea. Permissionless NFT markets, themselves implemented as smart contracts, enable users to sell and purchase NFTs in two ways: as a fixed-price sale or auction, in which competing bids are locked by the smart contract together with the NFT until a winner is found and the auction is cleared. With the admission of NFTs into popular culture, trade volumes on these markets have seen dramatic growth from a mere \$12m settled in September 2020, to volumes exceeding \$3.5b in September of the following year, a surge of over 29,060\% \footnote{https://dune.xyz/sophieqgu/NFT-Marketplaces}. 

Users typically connect to the permissionless markets through public-key cryptography capable of generating an arbitrary number of addresses \cite{jensen_2021}. As a consequence, user identities remain entirely pseudonymous in NFT markets, making the obfuscation of illicit practices challenging to prevent. As the unique properties of the Ethereum blockchain simplifies adversarial agents to hide in plain sight, we hypothesize that wash trading and strategic bidding amongst multiple addresses controlled by a single or colluding agent, may be a frequent occurrence. As there are no theoretical limits to the number of pseudonymous addresses a single agent can control, we conjecture that adversarial agents likely employ a mixture of manual trading and bots to trade NFTs between clusters of addresses in their control. This behaviour serves the strategic purpose of artificially inflating the trade volume of a given NFT, creating an impression of desirability to uninformed traders \cite{imisiker_2018}. The uninformed traders, looking for a great opportunity to buy a ‘hot’ NFT will interpret the transaction volume as an authentic expression of interest from other collectors and immediately place a bid or purchase the NFT at an artificially inflated price.

Furthermore, novel markets tend to be driven by a volatile search for suitable pricing models \cite{khuntia_2018}. This is undoubtedly the case for the blossoming crypto markets on which the current level of ‘irrational exuberance’ may result in inefficient markets given the presence of uninformed traders looking to strike gold. Adversarial market participants have been shown to exploit these conditions, primarily by employing strategic wash trading on centralized and decentralized central limit orderbook (CLOB) exchanges \cite{aloosh_direct_2019, cong_crypto_2020, pennec_wash_2021,  victor_detecting_2021}. 

Yet, the extent to which these or equivalent practices are being used on NFT markets, remains unclear. To fill the gap, in this paper we study activities between addresses participating in the NFT markets on Ethereum. We pursue the research question: ‘To what extent does wash trading occur in smart contract-based NFT markets on Ethereum, and to which extent does this practice distort prices?’. 
Conceptualizing trading patterns as a graph and proposing two detection algorithms, we identify 2.04\% as the lower bound of suspicious sale transactions that closely follow the general definition of wash trading.

%%%%%%%%%%%%%%%%%%%%%%%%%%%%%%%%%%%%%%%%%%%%%%%%%%%%%%%%%%%%%%%%%%%%%%%%%%%%%%%%

\section{Literature Review}
Wash trading is a well-known phenomenon in traditional financial markets and refers to the activity of repeatedly trading assets for the purpose of feeding misleading information to the market \cite{cao_2014}. Typically, one or more colluding agents conduct a set of trades, without taking market risks, that lead to no change of the initial position of the adversarial agents.
Most of the early academic publications on wash trading in financial markets focused on colluding investor behaviour (e.g. \cite{grinblatt_2004}). Cao et al. \cite{cao_2014} were among the first to analyse wash trading by specifying trading patterns. Later, they extended their study using directed graphs on order book data \cite{cao_detecting_2016}. The literature on the identification of wash trading patterns in the cryptocurrency markets primarily emphasizes CLOB models on decentralized exchanges \cite{victor_detecting_2021} and centralized exchanges, where wash trading practices have been shown to be especially prevalent\cite{aloosh_direct_2019, cong_crypto_2020, pennec_wash_2021}. Perhaps because the introduction of CFMMs has nearly eliminated the efficacy of wash trading in smart contract-based markets for fungible assets, research on NFTs tends to emphasize either market dynamics and pricing \cite{nadini_mapping_2021, dowling_is_2021} or the technical design considerations \cite{regner_nfts_2019, wang_non-fungible_2021}. Thus far, little academic research has examined market abuse in smart contract-based NFT markets \cite{das_2021}. 
%%%%%%%%%%%%%%%%%%%%%%%%%%%%%%%%%%%%%%%%%%%%%%%%%%%%%%%%%%%%%%%%%%%%%%%%%%%%%%%%

\section{Methodology}
The Ethereum blockchain is a type of permissionless ledger, in which all transactions and state changes introduced by smart contracts are replicated across all participating nodes in the network \cite{antonopoulos_2018}. This introduces a high level of integrity to the database, but simultaneously requires the pseudonymization, as anyone with access to the database would otherwise be able to view the balances of users on the network. In Ethereum, this problem is solved with public-key cryptography \cite{antonopoulos_2018}. Any user on the network can generate a public/private key pair, which can subsequently be used to generate an arbitrary number of addresses. This design presents a fascinating paradox from the perspective of identifying market abuse: Pseudonymous identities are essential in protecting the privacy of benevolent users but, at the same time, they allow adversarial agents to hide in plain sight. Yet, due to the strict ordering of transactions and unique properties of NFT markets, blockchain transaction data presents a powerful and unique opportunity for pattern detection utilizing graph-based algorithms \cite{victor_detecting_2021, chen_graph_2020, weber_2019} and address clustering \cite{victor_clustering_2020, harrigan_2016}.

\textbf{Data Aggregation and Cleaning:}
We collect transaction data on the 52 leading ERC721 NFT collections on the Ethereum blockchain by trading volume, covering a period between the 1st of January 2018 until 21st November 2021. The dataset contains 21,310,982 transactions of 3,572,483 NFTs conducted by 459,954 addresses. Collectively, the dataset represents \$6.9b of the \$12.3b total trading volume (49.5\%)\footnote{https://dune.xyz/sophieqgu/NFT-Marketplaces} on all NFT markets since the first block of the Ethereum blockchain\footnote{The authors will open-source scripts and data upon publication.}.

We capture all blockchain transactions related to the selected NFT collections via the OpenSea API. We parse the dataset by ‘sale’ events emitted by an NFT contract when it is transacted on a smart contract-based marketplace, indicating that a change of ownership has been recorded on the blockchain, and ‘transfer’ events indicating that the NFT has been transferred from one address to another. The dataset was subsequently enriched with (I) historic USD prices for settlements in crypto and stablecoins via the Coingecko API and (II) blockchain-specific data via the Etherscan services. (III) To maintain an accurate overview of the NFT markets in the dataset, we collected the deployment date of the four largest NFT on-chain markets manually (Foundation, OpenSea, Rarible, Superrare) and matched the deployment dates with the event emissions. 

It should be noted that the dataset collected for this analysis pertains only to operations conducted within the Ethereum Virtual Machine (EVM), meaning that we knowingly omit any ‘off-chain’ transactions or bidding patterns from the analysis. Finally, we pre-processed the dataset with standardized scripts, eliminating a very small fraction of transactions due to obvious technical errors or trades against exotic assets for which the price data tends to be inaccurate.

\begin{figure}
\vspace{-12pt}
\centering
\begin{minipage}{.49\textwidth}
 \centering
 \includegraphics[width=0.975\textwidth, height=4cm]{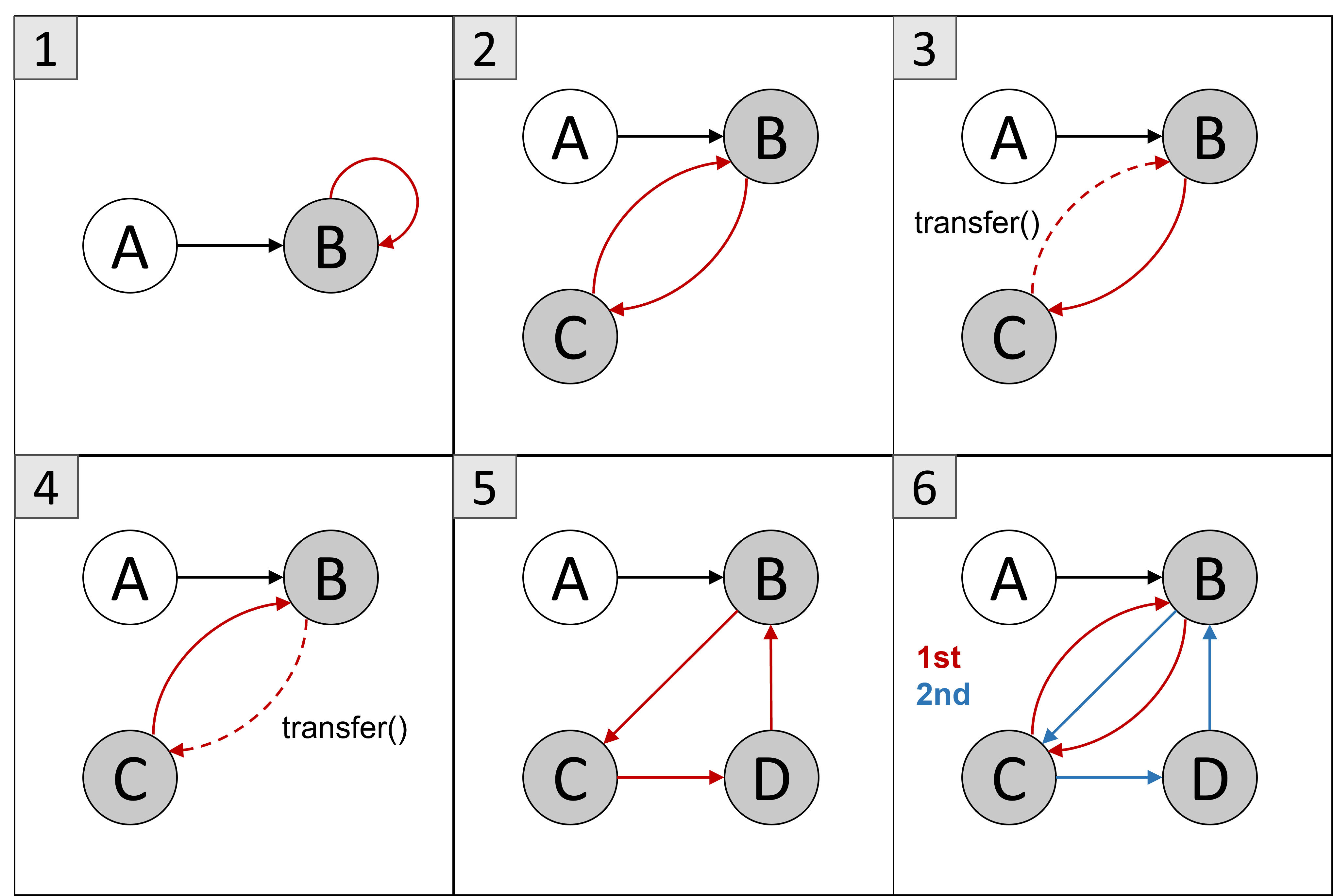}
 \caption{Detection of suspicious activity through closed cycles. We exclude cycles involving only ‘transfer’ events.}
 \label{fig:cycle}
\end{minipage}
\begin{minipage}{.49\textwidth}
 \centering
 \includegraphics[width=0.65\textwidth, height=4cm]{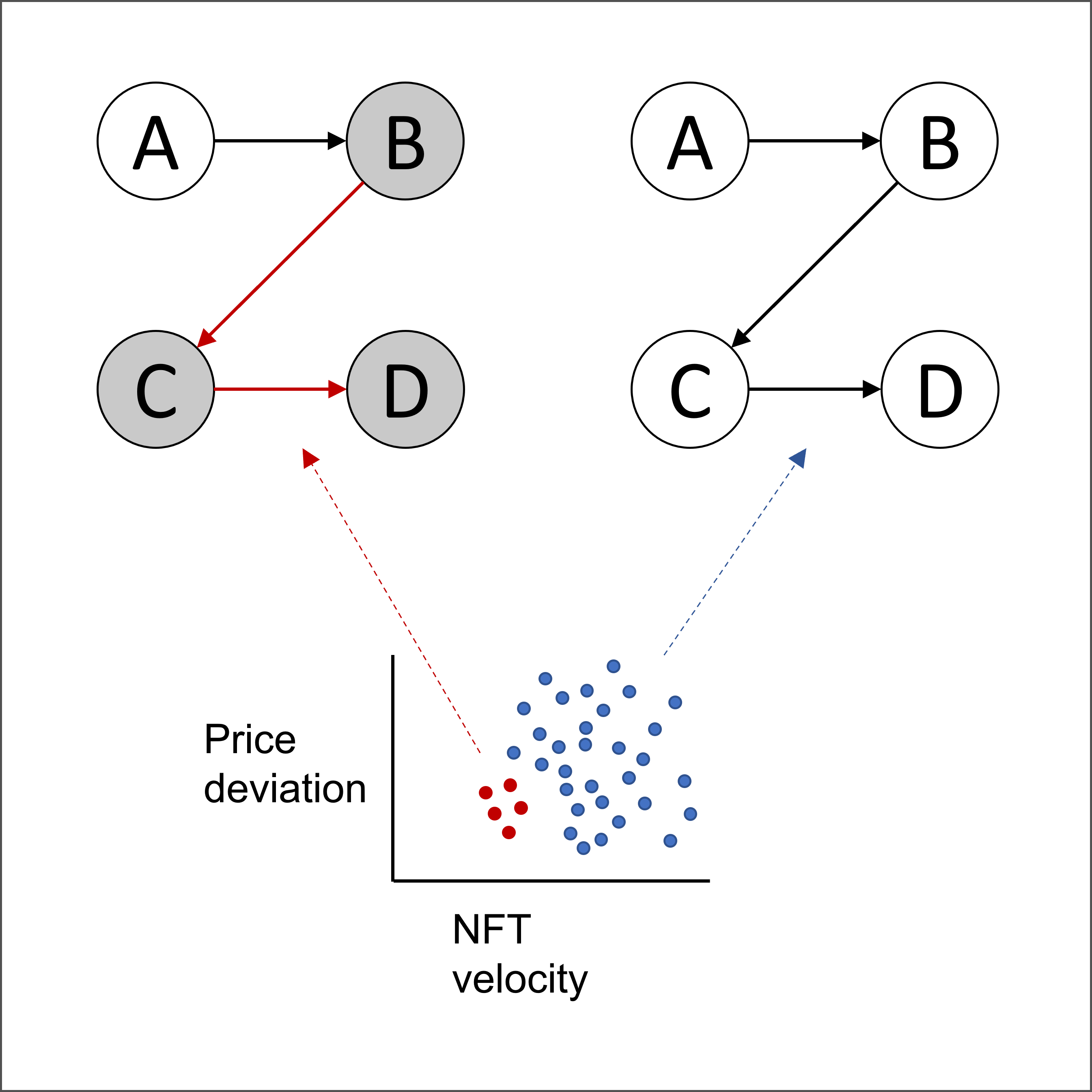}
 \caption{Detection of suspicious activity through a rapid sequence of transactions without taking market risk.}
 \label{fig:path}
\end{minipage}
\vspace{-12pt}
\end{figure}

\textbf{Building Transaction Graphs:}
In order to identify suspicious behaviour conforming with wash trading activity, we model the transaction history of each NFT as a directed multigraph $G_{nft}=(N, E)$, where $N$ is the set of addresses and $E$ is the set of transactions between addresses. The direction of the edges is given by the transaction flow from sender to receiver, identified by the transaction hash. The weight of the edges represents the USD price at the time of the transaction. 
This denotation is amenable to the identification of clusters in which a sequence of transactions leads to no apparent position change for any of the addresses involved. Topologically, these patterns form closed cycles. Utilizing Deep-First-Search-Algorithm \cite{tarajan_depth_1971} we identify closed cycles within the data set. We adjust and iterate the algorithm, using the temporal distance between transactions to detect sub-cycles. The proposed algorithm (Appendix) has a linear time complexity of ${\mathcal{O}((n+e)(c+1))}$ for \textbf{n} nodes (addresses), \textbf{e} edges (transactions) and \textbf{c} cycles \cite{johnson_finding_1975}.
Figure \ref{fig:cycle} illustrates a few examples of suspicious cyclic activities. Transactions $E$ belonging to a suspicious cycle are marked in red, and potentially colluding trading addresses $N$, are highlighted in grey. 

Example 1 is a self-directed transaction. Examples 2-4 illustrate variations of a cycle with two transactions, where solid lines represent ‘sales’ and dotted lines represent ‘transfers’. Example six depicts a complex graph where edges \{B, C\} and \{B, C, D\} form two sub-cycles distinct by time.

Further, we analyze path-like transaction patterns, as agents could actively avoid closed cycles. Informed by the definition of wash trading, we consider rapid trade sequences without exposure to market risk as potentially suspicious. Erring on the side of caution, we apply relatively strict thresholds. First, we define the transaction velocity for a sequence as the time elapsed from the initiation to the end. We delimit a rapid trade sequence below 12 hours. Second, we delimit the deviation in USD values, a proxy for market risk, in a sequence to a maximum of 5\% of the initial price. Combining both threshold flags 0.3\% of the sale transactions as mildly suspicious. Figure \ref{fig:path} showcases these path-like trade sets.

The proposed algorithms are highly applicable, in that NFT marketplaces deviate from conventional markets in several ways: First, as NFTs are uniquely identifiable by smart contract address and id, detection does not require volume matching required for fungible tokens (e.g. \cite{cao_detecting_2016, victor_detecting_2021}). Second, in contrast to other market designs such as CLOBs the seller can retain certain control over the opposing counterparty, making it potentially easier to conduct cyclical trades. Lastly, due to the transparency of the Ethereum blockchain, we can inspect trading behaviour at the account level without relying on statistical indicators \cite{pennec_wash_2021}.

%%%%%%%%%%%%%%%%%%%%%%%%%%%%%%%%%%%%%%%%%%%%%%%%%%%%%%%%%%%%%%%%%%%%%%%%%%%%%%%%

\section{Results}
The analysis flags a total of 3.93\% of the addresses as suspicious, indicating that these addresses might be controlled by single agents and used to conduct cyclical or sequential wash trading with NFTs. The flagged addresses processed 2.04\% of the total sale transactions, inflating the trading volume by \$149.5m or 2.17\% for the period. Of the 36,385 flagged sale transactions, 30,467 were conducted in clusters of cyclical patterns whereas 5,918 were conducted as a rapid sequence. The suspicious activity was executed with just 0.45\% of the NFTs in the dataset, indicating a high concentration of illicit activities around a few NFTs (Table \ref{tab:results_overview}).

\begin{table}
\vspace{-12pt}
\begin{center}
\caption{Overview of the results.}
\begin{tabular}{|l|l|l|l|}
\hline
 & Dataset & Identified & Percentage \\
\hline
Addresses & 459,954 & 18,117 & 3.93\% \\
Transactions & 1,779,380 & 36,385 (cyclic: 30,467 sequential: 5,918) & 2.04\% \\
Volume in \$ & 6.9 b & 149.5 m & 2.17\% \\
NFTs & 3,572,483 & 16,289 & 0.45\% \\
\hline
\end{tabular}
\label{tab:results_overview}
\end{center}
\vspace{-12pt}
\end{table}

While we identify suspicious activities in all NFT collections (Table \ref{tab:results_collections}, Appendix), the extent to which a collection is contaminated by flagged transactions ranges from 0.19\% to 60.93\%, indicating that adversarial agents tend to target specific collections for illicit practices. 
In general, we observe a predisposition for simple trading patterns. 60.6\% of the identified clusters are simple variations with two transactions (equivalent to examples 2-4 in Figure \ref{fig:cycle}). Complex variations of three (8.7\%) or more than three transactions are less common (30.7\%). However, we find no signs of self-directed trades.
Cyclical patterns are conducted at relatively rapid intervals. Figure 3 illustrates the elapsed time from the first to last transactions, with respect to the number of transactions involved. Overall, 48.1\% of the identified cycles happen within a single day. 13.2\% happen within one to seven days and 13.0\% are just below 30 days. Consequently, 74.3\% are conducted within 30 days, an important threshold under US regulation\footnote{https://www.cftc.gov/LawRegulation/CommodityExchangeAct/index.htm}. 

\begin{figure}
\vspace{-15pt}
\begin{center}
\includegraphics[width=0.975\textwidth, height=6cm]{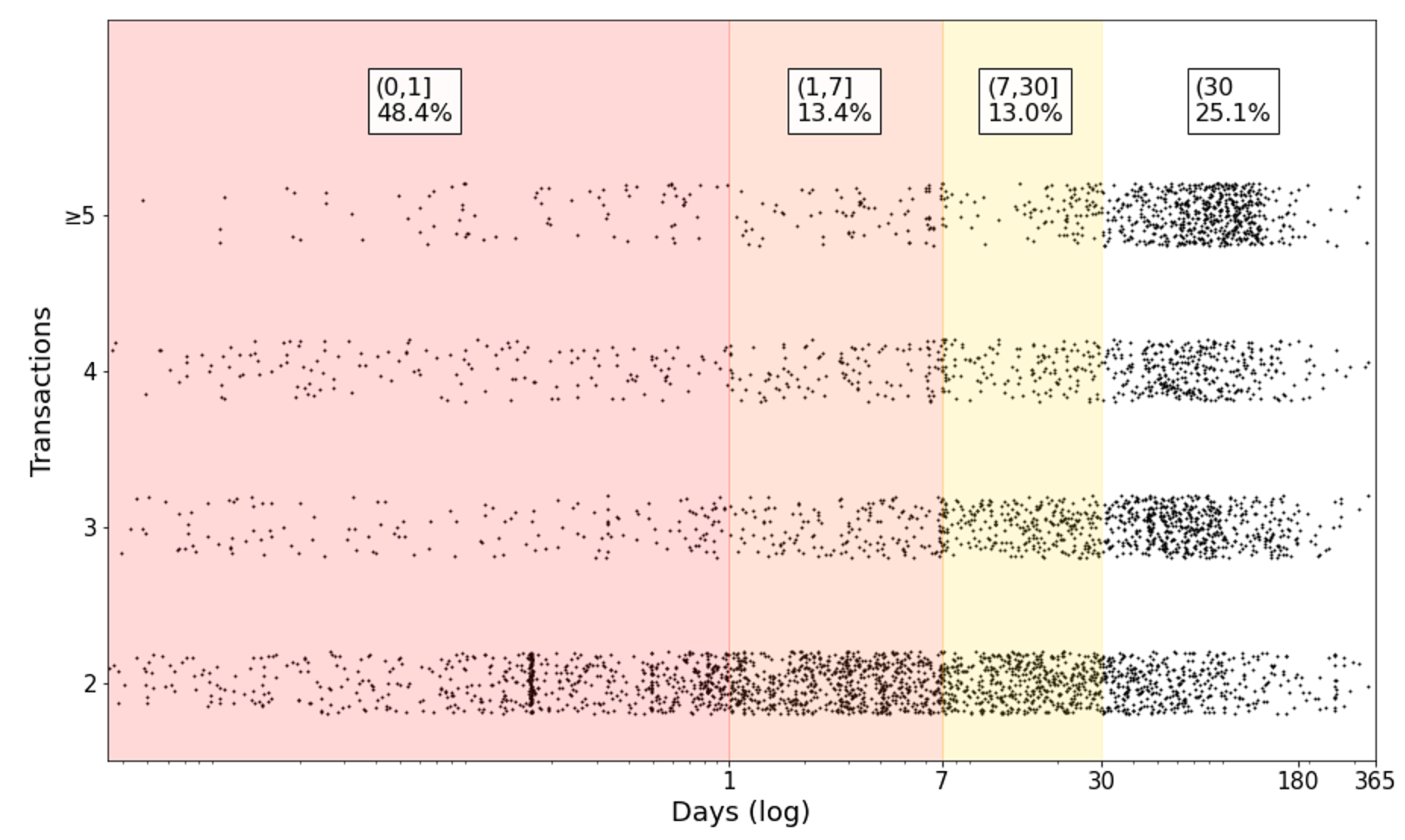}
\caption{Elapsed time to close a cycle with respect to the number of transactions involved. 48.1\% of the identified cycles happen within a single day.}
\end{center}
\label{fig:cycle_closed}
\vspace{-22pt}
\end{figure}

The identified 2-transactions variations have a median execution time of 4.2h (3-transactions: 54h), suggesting a preference for simple and fast patterns. We assume this to indicate that adversarial agents are not trading in an automated fashion, which would result in more complex patterns and execution times within a few minutes.
Rapid sequential trades, in which a NFT is moved fast between accounts without any market risk, contributed to only 5918 suspicious transactions, equivalent to 0.3\% of the transactions across all collections. 

\begin{figure}
\begin{center}
\includegraphics[width=0.8\textwidth]{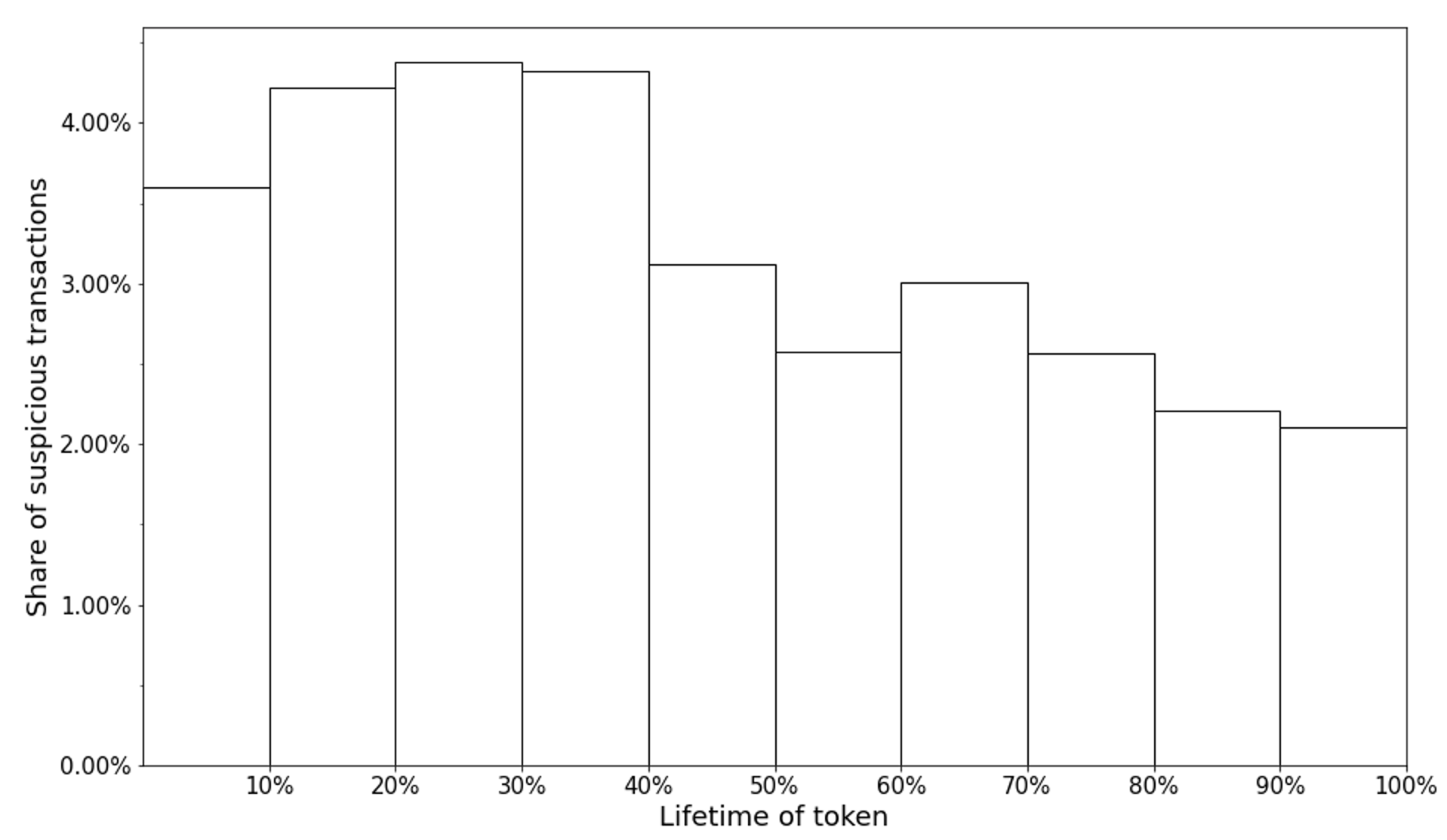}
\caption{Wash trading with respect to collections’ lifetime. In absolute terms, wash trading is the highest at the beginning of a collections’ lifetime, however this is also matched by a high amount of organic traffic.} 
\label{fig:washtrading_token_lifetime}
\end{center}
\vspace{-12pt}
\end{figure}

We investigate, at what point in the collections’ lifetime, suspicious activities occur. For each collection, we determine the mean suspicious activity starting with the creation date of the respective collection. Figure \ref{fig:washtrading_token_lifetime} (Appendix) shows a peak of suspicious activity in the first third of a collections’ lifetime, possibly in order to raise initial awareness to attract naive buyers. In absolute terms, wash trading is the highest at the beginning of a collections’ lifetime, however this is also matched by a high amount of organic traffic.
Increasing the price of an asset by faking activity is a central motivation for agents in conducting illicit trading \cite{imisiker_2018}. Analysing if the average price is inflated through wash trading practices, we find that the subsequent sale after a detected wash trade has, on average, an increased price of 30.53\%. However, a regression on panel data to measure the impact on the price, led to insignificant results for a majority of collections. Looking into external factors, we found that \textit{age} has a strong positive relationship with the price. While we expected an inverse relationship of the gas price, which impacts the costs per transaction, with the NFT price, we found the effect to be mixed in a majority of cases. We suspect these findings are influenced by the strong bull market, given the current overall positive public sentiment.

\begin{figure}
\begin{center}
\includegraphics[width=0.8\textwidth]{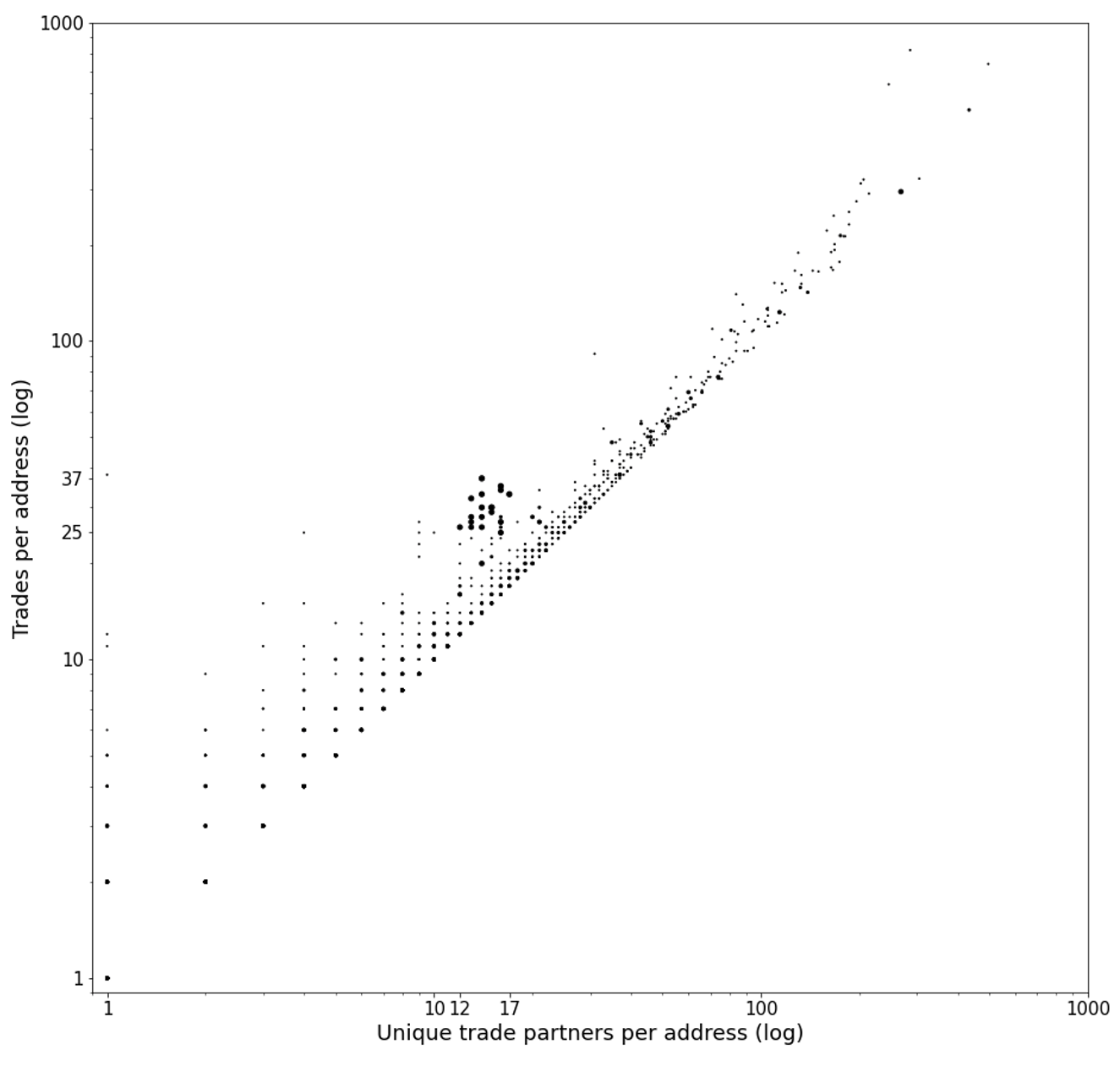}
\caption{Trades and unique trade partners. Each dot represents an address trading on NFT markets, positioned by the amount of trades and unique trade partners. The size of the dot depicts the number of empirically identified suspicious trades.} 
\label{fig:trade_partner}
\end{center}
\end{figure}

Finally, we explore the relationship between executed trades per address and unique trade partners per address. On NFT markets sellers can retain certain control over the opposing counterparty, thus a large number of trades with only few other addresses raises suspicion. Figure \ref{fig:trade_partner} (Appendix) visualizes this relationship, whereas addresses which conducted many trades with only a few other addresses would be tilted to the left. Each dot represents an address trading on NFT markets, positioned by the amount of trades and unique trade partners. The size of the dot depicts the number of empirically identified suspicious trades.
We find a cluster of suspicious addresses, conducting 25-37 trades with only 12-17 unique trade partners. Furthermore, in contrast to other markets, addresses have relatively few trades, again suggesting a low level of automated trading in this nascent market.

\section{Discussion}
Given the challenges in interpreting pseudonymous blockchain based data, this study has multiple limitations. First and foremost, it should be made clear that none of the findings presented in this paper present any conclusive evidence of criminal activities or malicious intent.
While we delimit a set of behaviours which we find unlikely to be conducted with benevolent intent, we leave it to the reader to assess the likelihood that flagged transactions constitute attempts at wash trading. Any, or all, of the flagged sequences may be erroneous but authentic transactions. Second, the decision of limiting our analysis to a specific subset of cyclical and sequential patterns may result in false negatives as sophisticated attempts at wash trading involving advanced address clusters over longer periods are not flagged by the analysis, at this point. Wash traders may me more careful and evade the analyzed heuristics. Similarly, the analysis pertains exclusively to on-chain transaction events emitted by the NFT contract and does not account for strategic bidding practices, which we suspect may be a popular methodology amongst adversarial agents. Because of these limitations, we hypothesize that the results presented here detect a lower bound for actual extent of adversarial behavior on NFT markets.

Should wash trading conducted according to the patterns explored in this paper increase over time, smart contract-based NFT platforms may consider the implementation of obligatory or voluntary identification initiatives. Alternatively, trading limitations on trading velocity, price deltas or counterparties can be implemented. In our sample self-directed trades have been non-existent, with indicates successful countmeasures at smart contract or frontend level.  
Nevertheless, any such attempt at introducing restrictions or limitations may stifle organic market activity and will inevitably create a cat and mouse game, as developers and wash traders race to identify and create increasingly sophisticated patterns.

More subtle countermeasures fostering the supervision of NFT markets are the expansion of NFT standards beyond ERC721 and ERC1155, as well as increased data ubiquity. Decentralized NFT markets are transparent, however NFT data is very diverse and difficult to retrieve. Fees potentially play a big role in preventing wash trading, as long as the rewards or incentives are less than the cost-of-attack. Fees on NFT markets are substantial. Fraudulent agents are less likely to perform a wash trade if they are losing several percentages with every transaction. Admittedly, this does neither stop marketplaces itself to perform wash trades nor prevents private offset agreements between the trader and a marketplace.

Even so, with \$149.5m and a median of 2.04\% suspicious sale transactions we argue that wash trading may be less common than what industry observers have previously estimated \cite{coindesk_2020, bloomberg_2021}.

\section{Conclusion}
We identify what we believe may serve as a lower bound estimation for suspicious trading behaviour on NFT markets, following the definition of wash trading: sets of trades between collusive addresses, without taking market risk, that lead to no change in the individual position of the participating addresses. Our findings indicate that (I) adversarial agents exhibit a clear preference towards fast and simple cyclical patterns, (II) the level of suspicious activity varies significantly across NFT collections, (III) illicit activity could still be done in a manual fashion, and (IV) the activities do not necessarily produce the intended price impact, as other exogenous factors such as \textit{age} and \textit{sentiment} are more relevant to price discovery.

As a theoretical contribution, we add descriptive knowledge to an emerging field of research where scientific studies are scarce. We contribute to the growing literature on the identification of illicit market behavior in centralized and decentralized crypto markets, by conducting the first in-depth examination of NFT wash trading on the Ethereum blockchain. We contribute empirical statistics of fraudulent behaviour and a set of suspicious transaction graphs to foster the understanding of wash trading in increasingly financialized NFT markets. The valuable insights we generate for practitioners are twofold: First, we provide valuable insights to prevent collectors from buying NFTs that are potentially inflated by wash trading. Second, we discuss practical countermeasures increasing the standards for the wider NFT ecosystem.
Further research opportunities are manifold and include studying NFT markets on other blockchains as well as researching the correlation between suspicious behaviour and sentiment data.

%%%%%%%%%%%%%%%%%%%%%%%%%%%%%%%%%%%%%%%%%%%%%%%%%%%%%%%%%%%%%%%%%%%%%%%%%%%%%%%%
\pagebreak

\printbibliography

\begin{center}
\includegraphics[width=0.8\textwidth]{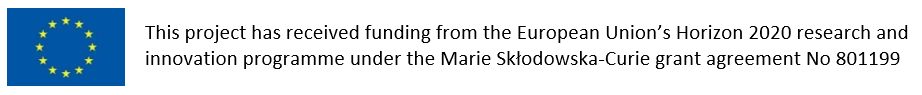}
\end{center}

%%%%%%%%%%%%%%%%%%%%%%%%%%%%%%%%%%%%%%%%%%%%%%%%%%%%%%%%%%%%%%%%%%%%%%%%%%%%%%%%
\pagebreak

\section{Appendix}
\textbf{Algorithm}
\begin{algorithm}
\caption{The detection algorithm}
\label{alg:detection}
\begin{algorithmic}[1]
\State Input: $T$ timestamped blockchain transactions
\State $L \gets$ empty list of $cycles$

\For{$nft \in T$}
 \State $G_{nft} \gets (N, E)$
 \State $G_{nft} \gets$ identifier, weight
 \State label $n \in N$ as discovered
 \For{all directed $E$ of $n$}
 \State test for adjacent edges $m$
 \If{$m$ is not labeled as discovered}
 \State continue
 \Else
 \State $L \gets cycle$
 \State $G_{nft}* \gets G_{nft} - E$
 \State break and recurse
 \EndIf
 \EndFor
\EndFor
\State return $L$
\end{algorithmic}
\end{algorithm}

\begin{table}
\begin{center}
\caption{Results for each collection. Column (A) is the share of suspicious addresses, (B) the share of suspicious transactions, (C1) represents the total volume flagged denominated in USD, (C2) the share of the flagged volume and (D) the share of suspicious NFTs}
\begin{filecontents*}{statistics.csv}
,Collection,Smart contract,Created,Size,(A),(B),(C1),(C2) ,(D)
1,0n1-force,0x3bf2...5e9d,08/2021,7776,4.2,1.6,1732469,1.2,1.7
2,acclimatedmooncats,0xc3f7...5e69,04/2021,18412,2.8,1.3,357341,0.8,0.5
3,adam-bomb-squad,0x7ab2...78c5,08/2021,25000,1.3,0.5,206478,0.5,0.2
4,autoglyphs,0xd4e4...7782,04/2019,512,2,0.7,206032,0.5,0.4
5,axie,0xf5b0...cb8d,04/2018,243000,1.6,0.3,285055,1.4,0.1
6,bored-ape-kennel-club,0xba30...5623,06/2021,10000,3.2,1.3,1423011,1.3,0.9
7,boredapeyachtclub,0xbc4c...f13d,04/2021,10000,4,1.4,11308109,1.4,1.6
8,collectvox,0xad9f...d34c,08/2021,8888,1.7,0.6,138121,0.4,0.5
9,cool-cats,0x1a92...050c,06/2021,9933,4,1.3,2299930,1.2,1.4
10,creature-world-collection,0xc92c...aafc,08/2021,10000,2.6,1,862273,0.9,0.9
11,cryptoadz-by-gremplin,0x1cb1...49c6,09/2021,7025,5.4,2.1,2964440,1.7,2.2
12,cryptokitties,0x0601...266d,01/2018,2009725,0.5,2.3,421288,1.6,0
13,cryptopunks,0xb47e...3bbb,01/2018,10000,10.2,4.3,53892061,2.4,3.7
14,cryptovoxels,0x7998...cf0c,06/2018,6210,1.5,0.4,54925,0.2,0.4
15,cyberkongz,0x57a2...4f37,04/2021,4147,3.1,1.6,824246,0.7,1
16,cyberkongz-vx,0x7ea3...7c8b,08/2021,14334,2.1,0.7,326717,0.5,0.4
17,deadfellaz,0x2aca...a17b,08/2021,10000,1.4,0.6,175900,0.6,0.5
18,decentraland,0xf87e...5d4d,04/2018,92598,9.2,9.7,3312118,7.5,0
19,doodles,0x8a90...992e,10/2021,10000,2.4,1.1,757788,0.8,0.7
20,fluf-world,0xccc4...a68d,08/2021,10000,2.8,1,434361,0.8,0.7
21,foundation,0x3b3e...5405,01/2021,103251,9.6,7.8,383853,11.2,0
22,galacticapes,0x12d2...4d14,09/2021,10000,3,1,540459,1,0.8
23,galaxyeggs,0xa081...7c48,09/2021,9999,2.8,1.1,577130,1.3,0.9
24,hashmasks,0xc2c7...6928,01/2021,16384,4.6,2.3,1493358,1.8,1.5
25,jungle-freaks-by-trosley,0x7e6b...4de0,10/2021,10000,3,1.4,887701,1.3,1.2
26,koala-intelligence-agency,0x3f5f...6360,08/2021,10000,2.5,1,393869,1,0.8
27,lazy-lions,0x8943...37e0,08/2021,10080,1.4,0.5,358509,0.6,0.5
28,lootproject,0xff9c...13d7,08/2021,7779,6.6,2.5,9458899,3.6,1.6
29,lostpoets,0xa720...f466,09/2021,27515,0.4,0.2,37669,0.1,0
30,meebits,0x7bd2...6bc7,05/2021,20000,2.1,0.8,1322740,0.6,0.3
31,mekaverse,0x9a53...ca8f,10/2021,8888,2.5,1.5,2173946,1.4,0.9
32,mutant-ape-yacht-club,0x60e4...a7c6,08/2021,30003,3.4,2,7253896,1.7,0.6
33,mutantcats,0xaadb...e46a,10/2021,10000,1.3,0.4,256031,0.5,0.4
34,pudgypenguins,0xbd35...2cf8,07/2021,8888,3.2,1.1,1764016,1.3,1.5
35,punks-comic,0x5ab2...c948,05/2021,10000,67,56.4,13674620,38.7,19.1
36,rari721,0x60f8...5ee5,05/2020,155346,6.4,5.7,2905148,14,0.1
37,rumble-kong-league,0xef01...909a,07/2021,10000,1.1,0.4,151521,0.5,0.2
38,sadgirlsbar,0x335e...b2d8,08/2021,10000,1.3,0.6,17836,0.4,0.4
39,sandbox,0x50f5...6d4a,12/2019,166464,0.9,0.5,200218,0.2,0.1
40,sneaky-vampire-syndicate,0x219b...2539,09/2021,8888,3.8,1.8,928516,1.4,1.2
41,somnium-space,0x595f...a0fa,10/2019,5025,1.3,0.5,244005,1.8,0.4
42,sorare,0x629a...6205,07/2019,329383,13.4,2.3,489009,0.7,0.7
43,supducks,0x3fe1...cbc5,07/2021,10000,2.8,1.1,544447,1,0.9
44,superrare1,0x41a3...850d,04/2018,4436,3.9,1.2,92844,0.2,0.6
45,superrare2,0xb932...b9e0,09/2019,4436,3.2,0.9,419872,0.7,2
46,the-doge-pound,0xf4ee...d043,07/2021,10000,2,0.8,725137,0.9,0.7
47,the-sevens-official,0xf497...187a,09/2021,7000,3.4,1.9,878788,2.1,1.3
48,thehumanoids,0x3a50...0edd,09/2021,10000,1.7,0.7,261715,0.7,0.6
49,tom-sachs-rockets,0x1159...5d26,08/2021,2000,76.9,60.9,16907477,58.6,54.3
50,veefriends,0xa3ae...beeb,05/2021,10255,1.8,1.4,662073,0.7,0.3
51,world-of-women-nft,0xe785...5330,07/2021,10000,1.6,0.5,539001,0.7,0.5
52,wrapped-mooncats,0x7c40...3572,03/2021,8903,15,6.1,981392,6.5,3.8
\end{filecontents*}
\csvautotabular{statistics.csv}
\label{tab:results_collections}
\end{center}
\end{table}

\end{document}